\documentclass[%
reprint,
superscriptaddress,
showpacs,
amsmath,amssymb,
prc,
floatfix, ]%
{revtex4-1}

\usepackage{color}

\usepackage{graphicx}% Include figure files
\usepackage{dcolumn}% Align table columns on decimal point
\usepackage{bm}% bold math
\usepackage[dvipdfm,bookmarks=true,colorlinks,%
            citecolor=blue,linkcolor=blue,hypertex, %
            breaklinks=true]{hyperref}

\begin{document}

%\begin{CJK*}{GBK}{}

\title{Potential Energy Surfaces of Actinide Nuclei from a Multi-dimensional 
       Constraint Covariant Density Functional Theory: 
       Barrier Heights and Saddle Point Shapes}

\author{Bing-Nan Lu}% (ÂÀ±þéª)}%
%\email{bnlu@itp.ac.cn}
 \affiliation{Key Laboratory of Frontiers in Theoretical Physics,
              Institute of Theoretical Physics, Chinese Academy of Sciences,
              Beijing 100190, China}
\author{En-Guang Zhao}% (ÕÔ¶÷¹ã)}%
%\email{egzhao@mail.itp.ac.cn}
 \affiliation{Key Laboratory of Frontiers in Theoretical Physics,
              Institute of Theoretical Physics, Chinese Academy of Sciences,
              Beijing 100190, China}
 \affiliation{Center of Theoretical Nuclear Physics, National Laboratory
              of Heavy Ion Accelerator, Lanzhou 730000, China}
 \affiliation{School of Physics, Peking University,
              Beijing 100871, China}
\author{Shan-Gui Zhou}% (ÖÜÉƹó)}%
 \email{sgzhou@itp.ac.cn}
%\homepage{http://www.itp.ac.cn/~sgzhou}
 \affiliation{Key Laboratory of Frontiers in Theoretical Physics,
              Institute of Theoretical Physics, Chinese Academy of Sciences,
              Beijing 100190, China}
 \affiliation{Center of Theoretical Nuclear Physics, National Laboratory
              of Heavy Ion Accelerator, Lanzhou 730000, China}

\date{\today}

\begin{abstract}
For the first time the potential energy surfaces of actinide nuclei in 
the $(\beta_{20}, \beta_{22}, \beta_{30})$ deformation space are obtained 
from a multi-dimensional constrained covariant density functional theory.
With this newly developed theory we are able to explore the importance of 
the triaxial and octupole shapes simultaneously along the whole fission path.
It is found that besides the octupole deformation, the triaxiality also 
plays an important role upon the second fission barriers.
The outer barrier as well as the inner barrier are lowered by the 
triaxial deformation compared with axially symmetric results.
This lowering effect for the reflection asymmetric outer barrier is 
0.5 $\sim$ 1~MeV, accounting for $10 \sim 20\%$ of the barrier height.
With the inclusion of the triaxial deformation, a good agreement 
with the data for the outer barriers of actinide nuclei is achieved.
\end{abstract}

%21.60.Jz       Nuclear Density Functional Theory and extensions
%               (includes Hartree-Fock and random-phase approximations)
%24.75.+i       General properties of fission
%25.85.-w       Fission reactions
%27.90.+b       A ¡Ý 220

\maketitle

%\end{CJK*}

Since the first interpretation of the nuclear fission by the barrier 
penetration~\cite{Bohr1939_PR056-426},
it has been a difficult task to describe this phenomenon theoretically.
For example, in order to study the fission problem, one should first
have very accurate information about the fission barrier;
a 1~MeV difference in the fission barrier could result in several orders
of magnitude difference in the fission half-life.
Particularly, to explore the island of stability of superheavy nuclei 
(SHN)~\cite{Hofmann2000_RMP72-733, Morita2004_JPSJ73-2593, 
Oganessian2007_JPG34-R165, *Oganessian2010_PRL104-142502},
it is more and more desirable to have accurate predictions of fission
barriers of SHN~\cite{Moeller2009_PRC79-064304, Pei2009_PRL102-192501, 
Xia2011_SciChinaPAM54S1-109, Liu2011_PRC84-031602}.

Nowadays three types of models are used for calculating fission barriers.
During a long period the majority of these works is based 
on the macroscopic-microscopic (MM) models~\cite{Moeller2001_Nature409-785,
Moeller2004_PRL92-072501, Moeller2009_PRC79-064304, Sobiczewski2007_PPNP58-292}.
The MM models make use of the Strutinsky shell correction method,
allowing fast calculations of multi-dimensional PES's
containing most of the important shape degrees of freedom.
Until now it is still an important candidate for large scale 
fission barrier calculations based on the examination of 
multi-dimensional PES's~\cite{Moeller2004_PRL92-072501}.
In recent years, the rapid development of the density functional 
theories (DFT) also makes it possible to calculate the fission barriers 
fully self-consistently~\cite{Samyn2005_PRC72-044316, 
Burvenich2004_PRC69-014307, Karatzikos2010_PLB689-72, Abusara2010_PRC82-044303}.
There are mainly two reasons to apply DFT's in the study of fission properties.
First, many new functional forms and effective interactions
are proposed with much better performances for the excited state 
as well as the ground state calculations~\cite{Bender2003_RMP75-121, 
Vretenar2005_PR409-101, Guo2007_PRC76-034317, Meng2006_PPNP57-470, Niksic2011_PPNP66-519}.
Fission barrier calculations are also helpful for developing these DFT's.
Second, in DFT's much more shape degrees of freedom can be included
self-consistently. 
For example, the symmetry-unrestricted Skyrme-Hartree-Fock-Bogoliubov model 
has been applied for the fission studies~\cite{Staszczak2009_PRC80-014309}.
Besides these two types of models, there exist also methods intending 
to combine the advantages of the MM and self-consistent models, such as 
the extended Thomas-Fermi method~\cite{Mamdouh1998_NPA644-389}.

The double-humped fission barriers of actinide nuclei can be used to 
benchmark the predictive power of theoretical models~\cite{Mirea2011_CEJP9-116, 
Girod1983_PRC27-2317, Blum1994_PLB323-262, Burvenich2004_PRC69-014307}.
Various shape degrees of freedom play important and different roles
in the occurrence and in determining the heights of the inner and outer barriers.
For examples, it has long been known from MM model calculations that 
the inner fission barrier is usually lowered when the triaxial deformation is allowed, 
while for the outer barrier the reflection asymmetric (RA) shape 
is favored~\cite{Pashkevich1969_NPA133-400, Moeller1970_PLB31-283}.
Later on these points were also revealed in the 
non-relativistic~\cite{Girod1983_PRC27-2317}
and relativistic~\cite{Rutz1995_NPA590-680, Abusara2010_PRC82-044303}
density functional calculations, respectively.
It is thus customary to consider only the triaxial and reflection 
symmetric (RS) shapes for the inner barrier and axially symmetric 
and RA shapes for the outer one~\cite{Egido2000_PRL85-1198, 
Bonneau2004_EPJA21-391, Moeller2009_PRC79-064304}. 
It has been pointed out that ``there is no reason for a fissioning 
actinide nucleus not to penetrate all symmetry-breaking shapes on its
way from the first (triaxial) to the second (mass-asymmetric) 
saddle''~\cite{Skalski1991_PRC43-140}.
The non-axial octupole deformations are considered in both the MM 
models~\cite{Jachimowicz2011_PRC83-054302} and the non-relativistic 
Hartree-Fock theories~\cite{Skalski2007_PRC76-044603}.
However, a multi-dimensional structure of PES's including both 
the triaxial and RA shape degrees of freedom has not been explored yet 
in the framework of covariant DFT.
In this paper, we will investigate the influence of the triaxiality 
and the octupole shape on the PES's all the way from the ground state 
to the fission configuration when both shape degrees of freedom 
are included simultaneously.
To this end, not only as many as self-consistent symmetries should 
be broken, but also multi-dimensional constraints are needed~\cite{Ring1980}.

To calculate the potential energy surfaces and fission barriers, in
this work we use the covariant density functional theory 
(CDFT)~\cite{Serot1986_ANP16-1, Ring1996_PPNP37-193, 
Vretenar2005_PR409-101, Meng2006_PPNP57-470, Niksic2011_PPNP66-519}.
By breaking not only the axial~\cite{Meng2006_PRC73-037303, Lu2011_PRC84-014328} 
but also the reflection symmetries~\cite{Geng2007_CPL24-1865}, 
we developed multi-dimensional constraint CDFT's 
in which the functional can be one of the following four forms: 
the meson exchange or point-coupling nucleon interactions combined with 
the non-linear or density-dependent couplings~\cite{Lu2011_in-prep}.
If not specified, the functional form of the point-coupling nucleon interaction 
with non-linear self energy terms and the parameter set 
PC-PK1~\cite{Zhao2010_PRC82-054319} are used in this work.

For the parametrization of the nuclear shape, we adopt the conventional
ansatz in mean field calculations,
\begin{equation}
 \beta_{\lambda \mu} = \frac{4\pi}{3 A R^\lambda} \langle Q_{\lambda \mu} \rangle
\end{equation}
where $Q_{\lambda \mu}$ is the mass multipole operators. 
When the axial and reflection symmetries are broken, the nuclear shape 
is invariant under the reversion of $x$ and $y$ axes. 
In other words, the intrinsic symmetry group is $V_{4}$ 
and all shape degrees of freedom 
$\beta_{\lambda\mu}$ with $\mu$ even, including the triaxial
($\mu \ne 0$) and octupole ($\lambda = 3$) deformations, are possible.
Irrespective with the self-consistent symmetries, the single particle wave 
functions and various densities are expanded on an axially deformed 
harmonic oscillator basis~\cite{Ring1997_CPC105-77, Lu2011_PRC84-014328}. 
In order to get a fast convergence of the results against the basis size, 
in the elongated direction more states are included in the basis. 
Following Warda et al.~\cite{Warda2002_PRC66-014310}, 
the basis is truncated as: $n_z/Q_z+(2n_\perp+|m|)/Q_\perp \le N_\mathrm{cut}$. 
Here $n_z, n_\perp$, and $m$ are quantum numbers characterizing each state in the basis,
$Q_z=\mathrm{MAX}(1, b_z/b_0)$, $Q_\perp = \mathrm{MAX} (1, b_\perp/b_0)$,
and $b_0$, $b_z$, and $b_\perp$ are the oscillator lengthes. 
The calculated binding energy of $^{240}$Pu at $\beta_{20}=1.3$ varies only about 
130 keV and 20 keV when $N_\mathrm{cut}$ increases from 16 to 18 and from 18 to 20. 
This means a good convergence and such a truncation scheme with $N_\mathrm{cut} = 16$ 
ensures a 0.2 MeV accuracy for the deformation range we are interested in. 
In the present work, $N_\mathrm{cut} = 16$ (20) is used in the triaxial (axial) calculations.
More details of the convergence study will be given in Ref.~\cite{ Lu2011_in-prep}.
The BCS approach is implemented in our model to take into account the
pairing effect.
Since it has been found that the BCS calculation with a constant pairing 
gap can not provide an adequate description of the fission 
barriers~\cite{Karatzikos2010_PLB689-72}, we use a delta force 
for the pairing interaction with a smooth 
cutoff~\cite{Zhao2010_PRC82-054319, Bender2000_EPJA8-59}.
\begin{figure}
\begin{centering}
\includegraphics[width=0.9\columnwidth]{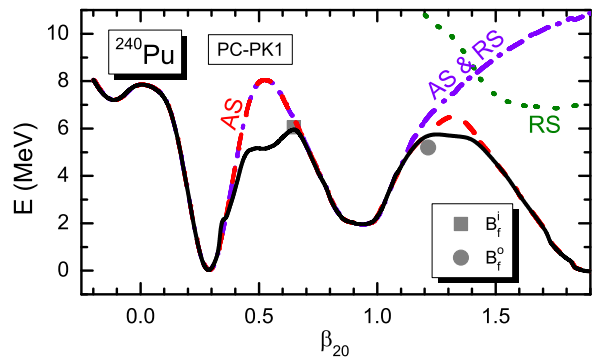}
\par\end{centering}
\caption{\label{Pic:PU240-1d}(Color online) 
Potential energy curves of $^{240}$Pu with various self-consistent symmetries imposed. 
The solid black curve represents the calculated fission path with $V_4$ symmetry imposed, 
the red dashed curve that with axial symmetry (AS) imposed, 
the green dotted curve that with reflection symmetry (RS) imposed, 
the violet dot-dashed line that with both symmetries (AS \& RS) imposed. 
The empirical inner (outer) barrier height $B_\mathrm{emp}$ is denoted by the grey square (circle). 
The energy is normalized with respect to the binding energy of the ground state. 
The parameter set used is PC-PK1.
}
\end{figure}

We performed one- (1-d), two- (2-d), and three-dimensional (3-d) constrained 
calculations for the actinide nucleus $^{240}$Pu. 
In Fig.~\ref{Pic:PU240-1d} we show the 1-d potential energy curves (PEC)
from an oblate shape with $\beta_{20}$ about $-0.2$ to the fission 
configuration with $\beta_{20}$ beyond 2.0 
which are obtained from calculations with different self-consistent symmetries 
imposed: the axial (AS) or triaxial (TS) symmetries combined with 
reflection symmetric (RS) or asymmetric cases. 
The importance of the triaxial deformation on the inner barrier and 
that of the octupole deformation on the outer barrier stressed by earlier 
studies~\cite{Girod1983_PRC27-2317, Abusara2010_PRC82-044303, Rutz1995_NPA590-680} 
are clearly seen here: 
The triaxial deformation reduces the inner barrier height by more than 2 MeV 
and results in a better agreement with the empirical datum;
the RA shape is favored beyond the fission isomer and lowers very much
the outer fission barrier. 
Besides these features, we observe for the first time that the outer
barrier is also considerably lowered by about 1 MeV when the triaxial 
deformation is allowed. 
Again, a better reproduction of the empirical barrier height can be seen for
the outer barrier.
We note that this feature can only be found when the axial and 
reflection symmetries are simultaneously broken.

\begin{figure}
\begin{centering}
\includegraphics[width=0.8\columnwidth]{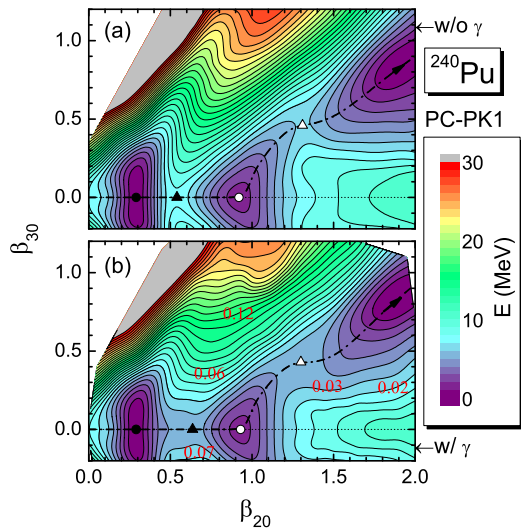}
\par\end{centering}
\caption{\label{Pic:PU240_2d}(Color online) 
Potential energy surfaces of $^{240}$Pu in the $(\beta_{20},\beta_{30})$ plane 
from calculations (a) without and (b) with the triaxial deformation included.
The energy is normalized with respect to the binding energy of the ground state. 
The numbers in (b) show the values of $\beta_{22}$ at these points.
The fission path is represented by a dash-dotted line. 
The ground state and fission isomer are denoted by full and open circles. 
The first and second saddle points are denoted by full and open triangles. 
The contour interval is 1 MeV. 
}
\end{figure}

How the PES of $^{240}$Pu becomes unstable against the triaxial distortion can 
be seen much more clear in Fig.~\ref{Pic:PU240_2d} in which we show 2-d
PES's from calculations without and with the triaxial deformation.
When the triaxial deformation is allowed, the binding energy of $^{240}$Pu 
assumes its lowest possible value at each $(\beta_{20},\beta_{30})$ point. 
At some points we get non-zero $\beta_{22}$ values. 
That is, non-axial solutions are favored at these points than the axial ones.
The triaxial deformation appears mainly in two regions in Fig.~\ref{Pic:PU240_2d}. 
One region starts from the first saddle point and extends roughly along 
the direction of the $\beta_{30}$ axis up to a very asymmetric shape
with $\beta_{30} \sim 1.0$.
In this region the values of $\beta_{22}$ are about $0.06\sim0.12$
corresponding to $\gamma \sim 10^{\circ}$. 
The energy, especially the inner barrier height, is lowered by about 2 MeV. 
The other region is around the outer barrier and the $\beta_{22}$ values
are about $0.02 \sim 0.03$ corresponding to $\gamma\sim$2$^{\circ}$. 
About 1 MeV is gained for the binding energy at the second saddle point
due to the triaxiality. 
In other regions, e.g., in the ground state and fission isomer valleys, 
only axially symmetric solutions are obtained.

\begin{figure}
\begin{centering}
\includegraphics[width=0.8\columnwidth]{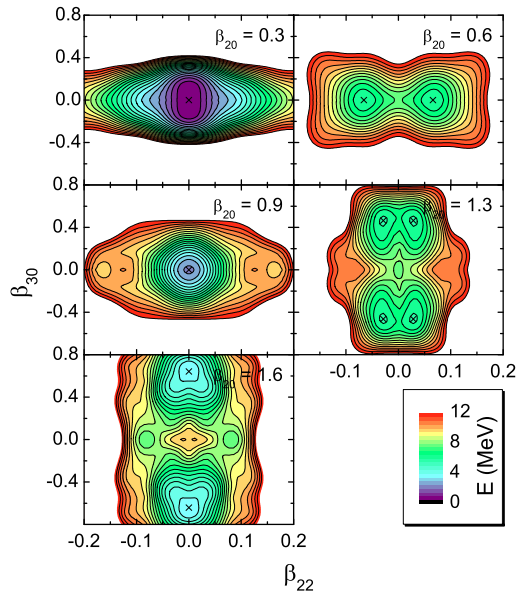}
\par\end{centering}
\caption{\label{Pic:PU240_3d}(Color online) 
Sections of the three-dimensional PES of $^{240}$Pu in the 
$(\beta_{22},\beta_{30})$ plane calculated at 
$\beta_{20}$= 0.3 (around the ground state),
0.6 (around the first saddle point), 0.9 (around the fission isomer), 
1.3 (around the second saddle point) and 1.6 (beyond the outer barrier), respectively. 
The energy is normalized with respect to the binding energy of the ground state. 
The contour interval is 0.5 MeV. 
Local minima are denoted by crosses.
}
\end{figure}

Next we examine the full 3-d PES of $^{240}$Pu obtained from the newly 
developed multi-dimensional constraint CDFT.
For simplicity, in Fig.~\ref{Pic:PU240_3d} are shown only five typical
sections of the 3-d PES of $^{240}$Pu in the 
$(\beta_{22},\beta_{30})$ plane calculated at 
$\beta_{20}$= 0.3 (around the ground state),
0.6 (around the first saddle point), 0.9 (around the fission isomer), 
1.3 (around the second saddle point) and 1.6 (beyond the outer barrier), respectively. 
Many conclusions can be drawn by examining these 3-d PES's. 
First, the ground state and the fission isomer are both axially and 
reflection symmetric as what is shown in the 1-d PEC and the 2-d PES.
But with the 3-d PES one can investigate the stability of $^{240}$Pu 
against the $\beta_{22}$ and $\beta_{30}$ deformations. 
One finds that the stiffness of the fission isomer is much larger than that 
of the ground state against both the $\beta_{22}$ and $\beta_{30}$ deformations. 
Second, while around the inner barrier the shape of $^{240}$Pu is triaxial
and reflection symmetric, the second saddle point which is close to 
$\beta_{20}=1.3$ appears as both triaxial and reflection asymmetric shape. 
Third, the triaxial distortion appears only on the top of the fission barriers. 

It has been pointed out that one may obtain spurious saddle points if only
a small number of shape degrees of freedom are constrained, see, e.g., 
Ref.~\cite{Moeller2009_PRC79-064304}.
That is, the calculated fission path may jump from one valley to another and
results in discontinuities in the lower-dimensional PES's; 
in some cases, a continuous path may even cross a higher saddle point. 
Although the spurious saddle points may not be excluded completely,
most of them can be avoided if (1) the obtained fission path keeps to be
continuous in the energy as well as the most important shape degrees of freedom 
and (2) the results are examined by higher-dimensional calculations. 
We have carefully checked the full 3-d PES and found that the fission path 
enters and exits the triaxial configuration rather smoothly, which tells 
that no sudden jump is found and the 1-d (with the $\beta_{20}$ deformation 
constrained and $\beta_{22}, \beta_{30}$ deformations imposed) and 2-d (with 
$\beta_{20}, \beta_{30}$ deformations constrained and the $\beta_{22}$ 
deformation imposed) calculations of the fission barriers may be well
justified for $^{240}$Pu.
It is clear that the continuity of the fission path found in a lower-dimensional
constraint calculation is a necessary but not sufficient condition 
for locating the correct saddle point.
In order to have a strictly definite conclusion, one certainly should carry out 
multi-dimensional constraint calculations with even higher-multipolarity
deformations included.

For the RS calculations, the triaxiality also lowers the fission path 
by a few MeV beyond the second saddle point. 
This point is illustrated by the dotted line in Fig.~\ref{Pic:PU240-1d} 
and the local minima with $\beta_{30} = 0.0$ in the $\beta_{20} = 1.6$ 
subfigures of Fig.~\ref{Pic:PU240_3d}. 
However, it is relatively unimportant, because the RA fission is still the
most favored one even when triaxiality is included.

Guided by the features found in the 1-d, 2-d, and 3-d PES's of $^{240}$Pu,
the fission barrier heights are extracted for even-even actinide nuclei
whose empirical values are recommended in RIPL-3 
(see Table XI in~\cite{Capote2009_NDS110-3107}).
The emphasis is put on the influence of the triaxial deformation on 
the two fission barriers. 

\begin{figure}
\begin{centering}
\includegraphics[width=0.7\columnwidth]{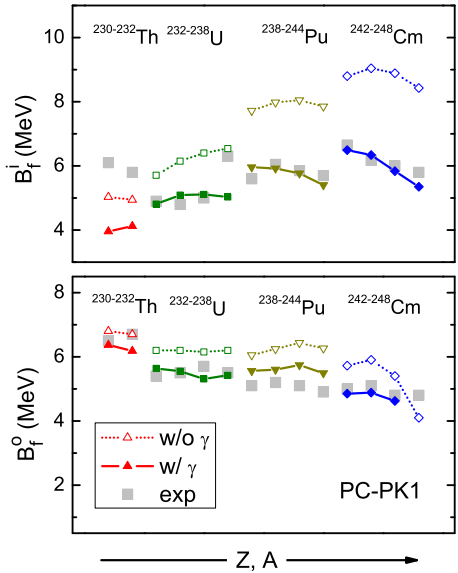}
\par\end{centering}
\caption{\label{fig:The-calculated-inner}(Color online) 
The inner ($B^\mathrm{i}_\mathrm{f}$) and outer ($B^\mathrm{o}_\mathrm{f}$) 
barrier heights of even-even actinide nuclei. 
The axial (triaxial) results are denoted by open (full) symbols.
The empirical values are taken from Ref.~\protect\cite{Capote2009_NDS110-3107}
and represented by grey squares.
}
\end{figure}

As it has been shown previously, around the inner barrier an actinide nucleus 
assumes triaxial and reflection symmetric shapes. 
Thus in order to obtain the inner fission barrier height we can safely 
make a one-dimensional constraint calculation with
the triaxial deformation allowed and the reflection symmetry imposed. 
In Fig.~\ref{fig:The-calculated-inner}(a) we present the calculated inner barrier 
heights $B^\mathrm{i}_\mathrm{f}$ and compare them with the empirical values.
It is seen that the triaxiality lowers the inner barrier heights
of these actinide nuclei by $1 \sim 4$ MeV 
as what has been shown in Ref.~\cite{Abusara2010_PRC82-044303}. 
In general the agreement of our calculation results with the empirical ones 
is very good with exceptions in the two thorium isotopes and $^{238}$U.
For $^{230}$Th and $^{232}$Th, the calculated inner barrier heights are
smaller by about 2 or 1 MeV than the empirical values depending on
whether the triaxial deformation is allowed or not. 
In these two nuclei, the outer barrier is higher than the inner one.
This may result in some uncertainties when determining empirically the height
of the inner barrier which is not the primary one~\cite{Samyn2005_PRC72-044316}. 
Similar results for $^{230}$Th and $^{232}$Th were obtained from the 
Skyrme-Hartree-Fock-Bogoliubov model~\cite{Samyn2005_PRC72-044316}
and very small inner barrier height was got for $^{232}$Th 
in Ref.~\cite{Abusara2010_PRC82-044303}. 
For $^{238}$U, $B^\mathrm{i}_\mathrm{f}$ from the axial calculation agrees
with the empirical value very well. The triaxiality reduces the barrier
height by about 1.5 MeV, thus bringing a discrepancy which was similar to
the result in Ref.~\cite{Abusara2010_PRC82-044303}.

\begin{figure}
\begin{centering}
\includegraphics[width=0.8\columnwidth]{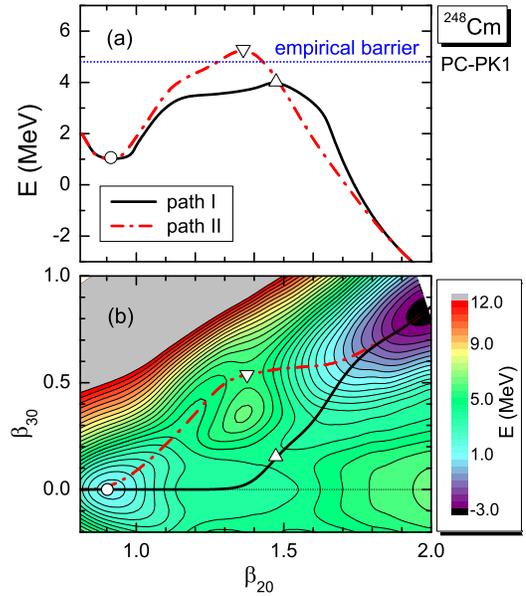}
\par\end{centering}
\caption{\label{fig:248Cm}(Color online) 
(a) One-dimensional potential energy curve $E \sim \beta_{20}$ and 
(b) two-dimensional potential energy surface $E \sim (\beta_{20},\beta_{30})$ 
of $^{248}$Cm the outer barrier region with the axial symmetry imposed in the calculation. 
In both figures, the energy is normalized with respect to the binding energy 
of the ground state. 
The fission path I/II is represented by full/dash dot lines
and the corresponding saddle point is denoted by up/down triangles. 
The fission isomer is denoted by open circle.
In (a), the empirical outer barrier height is depicted by dash-dotted line.
In (b), the contour interval is 0.5 MeV.
}
\end{figure}

To obtain the outer fission barrier height $B^\mathrm{o}_\mathrm{f}$, 
the situation becomes more complicated because more shape degrees of freedom 
have important influences around the outer fission barrier.
For example, the inclusion of the reflection asymmetric shape makes it
possible to have in the $(\beta_{20},\beta_{30})$ plane two or more competing 
fission paths with different octupole deformations.
In consequence, one often observes in the 1-d PEC two or more fission paths. 
This happens in $^{244,246}$Pu and $^{244,246,248}$Cm in the present study and 
we present a typical example in Fig.~\ref{fig:248Cm} for $^{248}$Cm. 
In this figure, one finds that there are two fission paths both in the 1-d 
$E \sim \beta_{20}$ curve and in the 2-d $E \sim (\beta_{20},\beta_{30})$ PES.
One path denoted by ``I'' favors shapes with larger octupole deformations and
the other denoted by ``II'' favors less RA shapes.
In such nuclei, it is not safe to perform a 1-d constraint calculation 
in order to get $B^\mathrm{o}_\mathrm{f}$.
Thus we first assume the axial symmetry and make a 2-d calculation in the 
$(\beta_{20},\beta_{30})$ plane from which we can approximately identify 
the lowest fission path $\beta_{30}^\mathrm{lowest}(\beta_{20})$ 
and the location of the second saddle point. 
Then along this fission path, we perform a 1-d $\beta_{20}$-constraint calculation 
with the triaxial and octupole deformations allowed. 
At each point with $\beta_{20}$, the initial deformations are taken as 
$\beta_{22}^\mathrm{ini.} = 0$ and 
$\beta_{30}^\mathrm{ini.} = \beta_{30}^\mathrm{lowest}(\beta_{20})$.
In this 1-d PEC, we can locate the second saddle point and extract the 
outer barrier height for each nucleus.

In the lower panel of Fig.~\ref{fig:The-calculated-inner} we show the
results of outer barrier heights $B^\mathrm{o}_\mathrm{f}$ and compare
them with empirical values. 
For most of the nuclei investigated here, the triaxiality lowers the outer barrier 
by 0.5 $\sim$ 1 MeV, accounting for about 10 $\sim$ 20\% of the barrier height.
One finds that our calculation with the triaxiality agrees well with 
the empirical values and the only exception is $^{248}$Cm.
From the calculation with the axial symmetry imposed, the outer barrier height 
of $^{248}$Cm is already smaller than the empirical value. 
The reason for this discrepancy may be related to that there are two 
possible fission paths beyond the first barrier, as seen in Fig.~\ref{fig:248Cm}. 
For the path I with a lower saddle point from which we get the outer fission
barrier height, the barrier is very wide and for the path II
with a higher saddle point, the barrier is relatively narrow. 
Therefore the empirical value of the outer fission barrier height may
not be easily extracted for the following two reasons: 
(i) There must be a strong competition between the two fission paths;
(ii) When the empirical value of the outer barrier height is evaluated, 
it is usually assumed that the second barrier is in an anti-parabolic
shape with a fixed and smaller width~\cite{Capote2009_NDS110-3107}. 

We also examined the parameter dependency of our results. 
The lowering effect of the triaxiality on the outer fission barrier
is also observed when parameter sets other than PC-PK1 are used.

In summary, a multi-dimensional constrained covariant density functional 
theory is developed which allows us to study the importance of the triaxial 
and octupole shapes simultaneously along the whole fission path.
The one-dimensional PEC $E \sim \beta_{20}$, two-dimensional
PES $E \sim (\beta_{20}, \beta_{30})$, and three-dimensional
PES $E \sim (\beta_{20}, \beta_{22}, \beta_{30})$ of actinide nuclei 
are shown and studied in details.
Both the triaxiality and the reflection asymmetry
plays crucial roles at and around the second saddle point.
The outer barrier as well as the inner barrier are lowered by the 
triaxial deformation compared with axially symmetric results.
For most of the nuclei investigated here, the triaxiality lowers the outer barrier 
by 0.5 $\sim$ 1 MeV, accounting for about 10 $\sim$ 20\% of the barrier height.
The calculated results of the outer barrier heights agree well with 
the empirical values. 

%\acknowledgements
Helpful discussions with 
Jie Meng, P. Ring, D. Vretenar, Xi-Zhen Wu, and Zhen-Hua Zhang 
are acknowledged.
This work has been supported by 
NSFC (Grant Nos. 10875157, 10975100, 10979066, 11175252, and 11120101005), 
MOST (973 Project 2007CB815000), 
and CAS (Grant Nos. KJCX2-EW-N01 and KJCX2-YW-N32). 
The computation of this work was supported by Supercomputing Center, CNIC of CAS.

%\bibliographystyle{apsrev4-1}
%\bibliography{/home/bnlu/Desktop/Works/Documents/Personal.Files/fission}
%\bibliography{../../../information/refs/JabRef/sgzhou}


\end{document}